# An efficient semi-Lagrangian algorithm for simulation of corona discharges: the position-state separation method

Lipeng Liu and Marley Becerra

*Abstract*—An efficient algorithm without flux correction for simulation of corona discharges is proposed. The algorithm referred to as the position-state separation method (POSS) is used to solve convection-dominated continuity equations commonly present in corona discharges modelling. The proposed solution method combines an Eulerian scheme for the solution of the convective acceleration, the diffusion and the reaction subproblems, and a Lagrangian scheme for the solution of the linear convection subproblem. Several classical numerical experiments in different dimensions and coordinate systems are conducted to demonstrate the excellent performance of POSS regarding low computational cost, robustness, and high-resolution. It is shown that the time complexity of the method when dealing with the convection of charged particles increases linearly with the number of unknowns. For the simulation of corona discharges where local electric fields do not change strongly in time, the time step of POSS could be much larger than the Courant–Friedrichs–Lewy (CFL) time step. These special features enable POSS to have great potential in modeling of corona discharges in long interelectrode gaps and for long simulation times.

## I. INTRODUCTION

Corona discharges occur when the electric field near an electrode becomes larger than the threshold electric field required to ionize the gas. Such situation arises when the characteristic size of the electrodes is much smaller than the interelectrode distance, such that a self-sustained, localized electrical discharge develops [1]. Corona discharges usually have two different modes: homogeneous glow corona and filamentary streamer discharges under different configurations and applied voltages [2].

Most simulation models of corona discharges in the literature follow the hydrodynamic approximation, where a set of continuity equations for the charged particles in the gas is solved simultaneously together with Poisson's equation. Continuity equations evaluate the variation of the density $\rho$ of charge particles (electrons, ions) in a gas under the influence of processes of convective bulk movement, diffusive motion and reaction terms. In general form, continuity equations are expressed as:

$$\frac{\partial \rho}{\partial t} + \nabla \cdot (\boldsymbol{u}\rho - D\nabla\rho) = f \qquad (1)$$

where $\boldsymbol{u}$ is the velocity vector of the considered particle and $D$ is the diffusion coefficient. The term $f$ defines the sources and sinks due to different processes such as impact ionization, attachment, photoionization, recombination, etc. Continuity equations in corona simulations are difficult to solve due to two main reasons. Firstly, the numerical time step is restricted by the solution of the continuity equation for electrons since they move more than two orders of magnitudes faster than ions. Secondly, discontinuities or steep gradients in $\rho$ or $\boldsymbol{u}$ can appear during the simulation when the transition to streamer discharges occurs. Under those conditions numerical methods may suffer from drawbacks such as artificial oscillations or excessive numerical diffusion [3].

Generally, the numerical methods for solving continuity equations can be divided into three groups based on the frame of reference: Eulerian, Lagrangian and mixed Eulerian-Lagrangian (also named as semi-Lagrangian). Let us now briefly describe the main features of the most common methods for solving the convection-dominated continuity equations for the simulation of corona discharges.

For most Eulerian methods, the 'flux-limiting' concept is generally used to avoid spurious oscillations occurring at discontinuities and shock fronts when high order spatial discretization schemes are used. One of the first methods to introduce this concept was the flux-corrected transport method (FCT) [4-6]. The FCT was originally proposed by Boris and Book for uniform meshes in one dimension and its extensions to high dimensions and non-uniform mesh were later suggested by Zalesak [7] and Morrow [8], respectively. Shortly afterwards, the finite-element FCT method (FEM-FCT) was introduced to handle irregular structures [9, 10]. FEM-FCT is based on Zalesak's FCT procedure dealing with element contributions, which was later modified by introducing the bilateral mass exchange between individual nodes [11]. However, Zalesak's strategy for flux-limiting is time consuming. Following the concept of flux-limiting, other high-resolution algorithms such as MUSCL (monotone upstream-centered schemes for conservation laws) scheme

This work was supported by the scholarship from China Scholarship Council (CSC) under Grant 201306160003.

The authors are with the School of Electrical and Engineering, KTH Royal Institute of Technology, SE-100 44, Stockholm, Sweden (e-mail: lipeng@kth.se and marley@kth.se).

[12], QUICK (quadratic upstream interpolation for convective kinetics) scheme [13] and TVB-DG (total-variation-bounded discontinuous Galerkin finite element method) [14-18] were also proposed. All of them can be viewed as different combinations of the finite element method (FEM) or the finite volume method (FVM) with different flux or slope limiters. However, the major disadvantage of most Eulerian, high-resolution schemes is that they are restricted by the Courant–Friedrichs–Lewy condition (CFL condition) and need flux correction, which is computationally expensive.

Purely Lagrangian methods are rarely used since severe problems occur when the discretization is performed on a moving Lagrangian mesh. Instead, mixed concepts that combine the advantages of Lagrangian and Eulerian methods are often used, which in essence follow the "exact-transport + projection" approach [19]. The basic idea is to use the Lagrangian method to calculate the exact transport of the unknown variables and then to project the obtained solution back onto an Eulerian mesh where the unknown variable is presented and other dependent variables are computed. The classical idea of semi-Lagrangian methods is to discretize the total derivate $\frac{D\rho}{Dt}$ (also called material derivative) in time instead of the partial derivative $\frac{\partial \rho}{\partial t}$ and most of them can only be applied to incompressible flows, i.e. with the restriction $\nabla \cdot \boldsymbol{u} = 0$ [19-24]. There are many variants of semi-Lagrangian methods in the literature. An example in the modelling of gas discharges is the particles-in-cell (PIC) method [22], which tracks individual particles in a continuous space and calculates the velocity field simultaneously on a stationary Eulerian mesh. Usually, super-particles need to be introduced to represent many real particles in order to make the simulations efficient, which introduces artificial numerical oscillations (discrete particle noise) in the solution [25, 26].

Summarizing, an improved numerical method suitable for corona discharges modelling should be:

- able to provide accurate, stable and non-negative solutions in the presence of shock fronts, discontinuities or steep gradients when filamentary streamer corona discharges take place;
- easy and fast, which means the algorithm is easy to implement and also computationally cheap to adapt multi-scale physical changes of corona discharges;
- easily implemented on unstructured meshes since the geometries where corona discharges present are often irregular such as point-to-plate or conductor-to-plate configurations;
- extendable to high dimensions and other coordinate systems for example cylindrical coordinate system which is frequently used when modelling corona discharges; and
- feasible for parallel computation to deal with huge unknowns in high-dimensional models.

In this paper, an efficient semi-Lagrangian method without flux correction for simulation of corona discharges is presented. The proposed approach, referred to as the position-state separation method (POSS), uses operator-splitting [27] to combine an Eulerian scheme for the solution of the convective acceleration, the diffusion and the reaction subproblems and a Lagrangian scheme for the solution of the linear convection subproblem. To handle arbitrary irregular geometries, the Eulerian solution is based on the Galerkin finite element method [28]. The Lagrangian subproblem is solved on an auxiliary mesh where each individual node during the simulation follows the flow only during the current time step, instead of using a moving mesh or a particle meshless concept. Even though high-resolution techniques can be used to project the solution from the auxiliary mesh to the Eulerian mesh, linear interpolation is here used as an excellent compromise between accuracy and computation time. Several classical tests are then performed to validate the proposed method, followed by an application example of corona discharges simulation.

The paper is organized as follows. The basic concept, the numerical implementation and the features of the method are introduced in Section II, III and IV, respectively. Section V is devoted to numerical experiments. Conclusions are drawn in the last section.

## II. MODEL AND METHOD FORMULATION

The target of corona discharges modelling is to estimate accurately the distribution of any particle at any time. In order to accomplish this target, the three essential elements of an unsteady particle flow – time, position and state – have to be calculated. When solving the continuity equation, difficulties arise since both the time and space discretizations of the density $\rho$ are mixed in one equation. To circumvent this, POSS uses the concept of splitting the transient solution of the position and the state of the density $\rho$ into two different numerical problems and to use the most suitable solution method for each of them.

Let us first consider the continuity equation in one dimension (1D) with a constant diffusion coefficient $D$. Then, equation (1) is rewritten as

$$\frac{\partial \rho}{\partial t} + u\frac{\partial \rho}{\partial x} = D\frac{\partial^2 \rho}{\partial x^2} - \rho\frac{\partial u}{\partial x} + f \qquad (2)$$

where $x \in [a, b], a < b$ is the computation domain discretized in space with an Eulerian mesh with $n$ nodes (hereinafter called reference mesh). By performing operator-splitting [27], equation (2) is divided into two different expressions such that its solution is approximated by the sequential solution of two separate subproblems:

i. the state subproblem given by

$$\frac{\partial \rho}{\partial t} = D\frac{\partial^2 \rho}{\partial x^2} - \rho\frac{\partial u}{\partial x} + f \qquad (3)$$

which deals with the variation of the variable $\rho$ due to diffusion, convective acceleration and the reaction terms. The solution of this equation provides the initial condition $\rho^*$ to the second subproblem during each time step.

ii. the position subproblem given by

$$\frac{\partial \rho}{\partial t} + u\frac{\partial \rho}{\partial x} = 0 \qquad (4)$$

which determines the transport of the variable $\rho$ by considering the linear convection only.

Since operator splitting is independent of the number of dimensions involved, this approach can be easily extended to two or three dimensions (2D/3D). Rewriting equation (3) and (4) in their multi-dimensional forms yields

$$\frac{\partial \rho}{\partial t} = D\Delta\rho - \rho\nabla\cdot\boldsymbol{u} + f \qquad (5)$$

$$\frac{\partial \rho}{\partial t} + \boldsymbol{u}\cdot\nabla\rho = 0 \qquad (6)$$

where the density $\rho$ is defined in a space variable $x$ belonging to the bounded domain $\Omega \in \Re^d$ ($d = 1, 2, 3$), and a velocity vector $\boldsymbol{u}$.

The state subproblem (5) is simple to solve with any Eulerian scheme in a reference mesh since the discretization of space and time is performed for different variables ($\rho$ and $\boldsymbol{u}$ respectively). Thus, equation (5) can be computed using a variety of numerical methods such as FVM or FEM (used in this paper). In turn, the solution to the position subproblem (6) is a difficult numerical problem in the presence of discontinuities or high gradients. In this paper, a scheme based on the method of characteristics is used to obtain the exact transport of the variable $\rho$ by using as initial value the solution of the state problem $\rho^*$ at each time step. Thus, equation (6) is solved by integrating the ordinary differential equation

$$\frac{\partial \boldsymbol{x}}{\partial t} = \boldsymbol{u} \qquad (7)$$

along the characteristic curves of the flow. In contrast to the other Lagrangian or meshless particles methods, individual parcels of fluid are not followed through time with a moving mesh or a system of particles. Instead, individual parcels of fluid located at the nodes of the reference mesh (used for the state subproblem) are tracked along their characteristic paths only during each time step. Then, the displacement of these parcels of fluid defines the nodes of an auxiliary mesh which is created from the reference mesh. One advantage of this approach is that the auxiliary mesh does not become excessively distorted (as a Lagrangian moving mesh) since it only "flickers" (depending on the variations of the local velocity vectors) around the corresponding nodes in the reference mesh. For this reason, the method does not require any mesh adaptivity or reconstruction algorithm, which is a computationally expensive step (especially in three dimensions). Compared with some particles methods, such as the particle transport method (PTM) [20, 21] which needs 'particle adaptivity' during each time step, the auxiliary mesh used here is also much easier to implement.

Once the solution to the position subproblem defines the location of the auxiliary mesh, the transported state of the density is projected back to the reference mesh. The projection is here done using linear interpolation. For a 1D mesh, the state at one position is obtained by linear interpolation from the two adjacent points. For a 2D or 3D mesh, the projection from the auxiliary mesh to reference mesh is implemented by Delaunay triangulation and linear interpolation with the scattered points (i.e. the nodes of the auxiliary mesh). As it will be discussed in Section IV, the Delaunay triangulation needs to be done only once since the element connectivity list does not change if the time step is properly selected.

POSS is essentially a modified semi-Lagrangian method that views convective acceleration ($\rho\nabla\cdot\boldsymbol{u}$) part as a reaction term to handle compressible flows. In order to further illustrate how POSS works, let us consider the 1D transport of charge particles drifting with constant velocity $u$, without diffusion and with a constant loss reaction term. Space is discretized with a reference mesh with the uniform spacing length $\Delta x$. Fig. 1 shows the sequential solution to the state and position subproblems after a simulation time step $\Delta t$ for an arbitrary profile of $\rho$ at the time $t_0$.

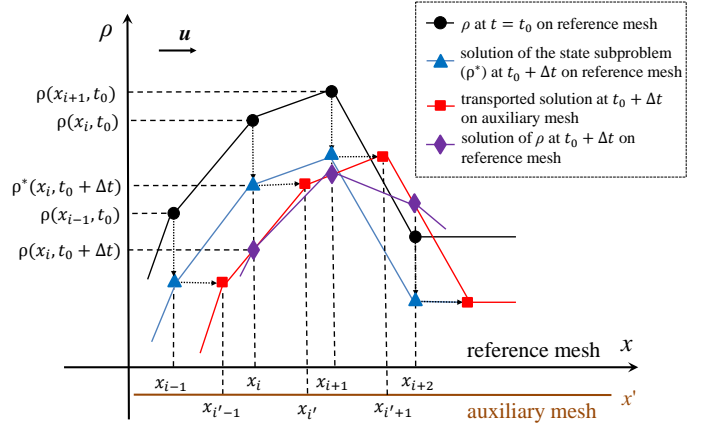

Fig. 1. Illustration of the steps of POSS to solve a 1D density profile during the time interval $[t_0, t_0 + \Delta t]$.

As it can be seen, the solution at each simulation time $t$ consists of three main steps:

- the estimation of the magnitude (i.e. the state) of the density $\rho$ in the reference mesh considering the convective acceleration as well as the diffusive and reaction terms;

- the exact transport of the updated state $\rho^*$ of the density such that the position of the solution in obtained in the auxiliary mesh; and

- the projection of the solution obtained in the auxiliary mesh back to the reference mesh.

III. NUMERICAL IMPLEMENTATION

Fig. 2 shows the flow chart of the numerical implementation of POSS according to the approach presented in the previous section. The details of each block in the flow chart are discussed in the following sub-sections. The main idea of the algorithm is described as follows. At first, the reference mesh is generated based on the computation region of the problem that is to be solved. The algorithm is continued step by step until the stop criterion is satisfied. During each simulation loop, the main calculation steps are executed. First, the state subproblem (5) is solved on the reference mesh. Second, the auxiliary mesh is generated by solving the position subproblem (6). Third, the solution after each step is obtained by projection of the solution in the auxiliary mesh to the reference mesh. These steps are followed by the adjustment of the solution according to the boundary conditions of the problem. The choice of time step $\Delta t$ will be discussed in details in Section IV.

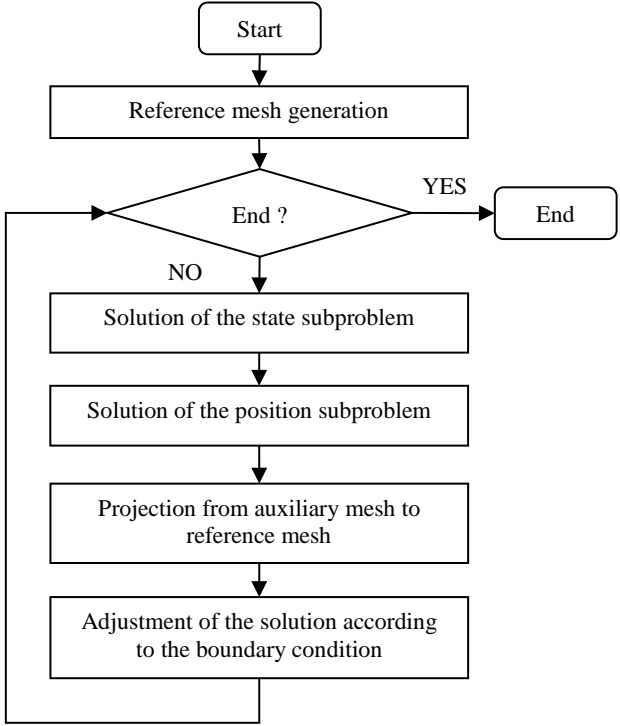

Fig. 2. The flow chart of POSS

*A. Mesh generation*

The POSS does not have any restriction regarding the type of elements used for meshing the reference mesh. In this paper, the Delaunay triangulation algorithm is used with triangular and tetrahedrons mesh elements in 2D and 3D simulations, respectively. The reference mesh can be unstructured where necessary. It is worth mentioning that the reference mesh can be 'stationary' throughout the calculations or it can be a 'moving' mesh when necessary as in the case of other adaptive moving-mesh Eulerian methods used to track the spatial variation of unknowns. In either case, the auxiliary mesh is updated at each time step from the location of the nodes in the corresponding reference mesh.

For sake of simplicity, linear basis functions are chosen to represent the unknowns. In this way, the calculation of the stiffness matrix and the consistent mass matrix for solving equation (5) using FEM can be computed analytically [29].

*B. Solution of the state and position subproblems*

The solution of the state subproblem on the reference mesh is here computed with the Galerkin finite element method [28]. If the diffusion term is not negligible, equation (5) is always solved using implicit schemes. Note that the boundary conditions have to be taken into account when solving the state subproblem (see [28] for more details).

Additionally, for convection-dominated flows when the diffusion term cannot be neglected, the calculation of the divergence of the velocity field (due to convective acceleration) in equation (5) is vital. Since linear basis functions are used, the divergence of the velocity at each element is constant. To obtain its value at each node, the divergence of velocity is computed using Galerkin FEM by

$$\boldsymbol{M}_c \nabla \cdot \boldsymbol{u} = \boldsymbol{f}_v \qquad (8)$$

where $\boldsymbol{M}_c$ is the global consistent matrix, $\nabla \cdot \boldsymbol{u}$ is the vector of velocity divergence and $\boldsymbol{f}_v$ is the contribution vector assembled by the constant value of each element. Due to the linear basis function, the local consistent matrix is easily computed analytically. Note that element contribution calculation can be done using parallel computation and $\boldsymbol{M}_c$ does not change for the reference mesh. Thus, equation (5) could be solved using the explicit method for convection-dominated continuity equations. For 1D problems, when $\boldsymbol{M}_c$ is usually not large, its inverse can be computed and stored at the beginning of the simulation. For 2D/3D problems, when the inverse of matrix $\boldsymbol{M}_c$ is usually impossible to be stored, the incomplete Cholesky conjugate gradient (ICCG) method [30] can be used to solve equation (8).

As mentioned before, the position subproblem could be solved by integrating equation (7) along the characteristic lines of the flow. This is an ordinary differential equation (ODE) which is easy to solve. First-order Euler method is used for ordinary flows. For the flows where the velocity field changes strongly in time and space, high order time discretization schemes have to be used to solve both the state and position subproblems in order to reach high-resolution. In addition, if the diffusion term is absent, the state subproblem could be viewed as an ODE which could be solved using Runge–Kutta (RK) method. While for the cases with diffusion term, high order backward differential formula (BDF) combined with Galerkin FEM is used.

*C. Projection and boundary conditions*

In order to describe more generally the projection step with POSS, a 2D computation region with inlet, outlet and walls is considered as an example in fig. 3. Both the reference and auxiliary meshes are also shown. The wall nodes are assumed to have zero velocity. As it can be seen, all the nodes except the inlet nodes are within the domain of the auxiliary mesh. For these nodes the projection is done by linear interpolation using the triangles of the auxiliary mesh that contain them. As it will be shown in Section IV, inversion of elements in the auxiliary mesh does not occur when a proper time step is chosen. For this reason it is not necessary to do Delaunay triangulation again as long as the reference mesh is unchanged. The information of the Delaunay triangulation when the reference mesh is generated can be stored and used when the projection is done. Extrapolation must be used for the projection of the inlet nodes since they are not within the domain of the auxiliary mesh. The linear extrapolation method is based on the least-squares approximation of the gradient at the inlet boundary. Linear interpolation is performed sequentially at each mesh node in the reference mesh in order to guarantee always a positive solution after the projection is done. Since each value of the nodes is independent of others, parallel computation could be applied to the projection step.

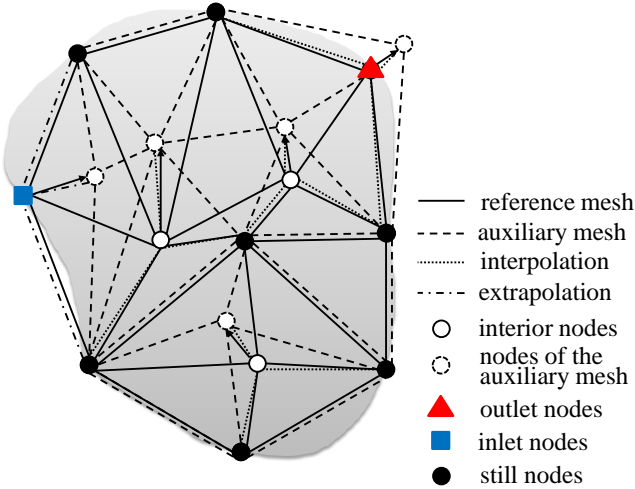

Fig. 3. Example of a two-dimensional reference mesh and the auxiliary mesh generated according to the local velocities

After the projection, the unknowns on the boundaries are adjusted according to the boundary conditions of the equation to be solved. For the Dirichlet condition, the value of unknowns at the boundary is set directly. For the Neumann condition no further action is needed since it has already been taken into consideration in the solution of the state subproblem.

## IV. FEATURES OF THE METHOD

### A. Stability

Let us take the case of one dimension as a first example. Since equation (4) is unconditionally stable, the only limitation comes from equation (3) if it is solved explicitly. In order to obtain a stable solution, it should be guaranteed that there is no inversion of any mesh element of the auxiliary mesh at any time. This means that the time step $\Delta t$ at each time of the simulation is restricted by the following condition:

$$\Delta t < \min_{1 \leq i \leq n-1} \left( \left| \frac{\Delta x_{i,i+1}}{\Delta u_{i,i+1}} \right| \right) \quad (9)$$

where $n$ denotes the total number of nodes, $\Delta x_{i,i+1}$ and $\Delta u_{i,i+1}$ are the space and velocity difference between the $i^{\text{th}}$ and $(i+1)^{\text{th}}$ nodes, respectively. This condition has the concrete physical meaning that the time step should be chosen short enough to guarantee that no parcel of the fluid transported in the position subproblem can overrun others. In this case, the adjacent nodes of each node of the auxiliary mesh do not change. The condition defines a maximum possible time step which is hereinafter referred to as the upper bound of stability $\Delta t_{UBS}$.

If we put this condition into equation (4), also expressed in the discretization form as $\frac{\Delta \rho}{\Delta t} = -\rho \frac{\Delta u}{\Delta x}$, the mathematic meaning will be derived: $\Delta \rho < -\rho$. It means that the decrement of the variable $\rho$ due to convection during each step should not exceed its absolute value. Thus, the variable $\rho$ cannot become negative if condition (9) is satisfied, making the POSS method be positive preserving. Following this idea, in multi-dimensions, $\Delta t_{UBS}$ can be expressed as:

$$\Delta t_{UBS}(t) = \min_{1 \leq i \leq n} \left( \frac{1}{|\nabla \cdot \boldsymbol{u_i}|} \right) \quad (10)$$

where $\nabla \cdot \boldsymbol{v_i}$ denotes the divergence of the velocity field.

### B. Accuracy

The error analysis in [31] shows that the overall error ($\varepsilon$) of semi-Lagrangian methods can be expressed as:

$$\varepsilon \sim \mathcal{O}\left( (\Delta t)^k + \frac{(\Delta x)^{P+1}}{\Delta t} \right) \quad (11)$$

where $k$ and $P$ are the order of backward time integration and the order of the interpolation, respectively [24]. $\Delta t$ and $\Delta x$ refer to time and space stepping. Expression (11) shows that the error of a semi-Lagrangian method is not monotonic with respect to neither $\Delta t$ nor $\Delta x$. In order to illustrate the error of POSS, let us consider the 1D non-diffusive transport of a unitary square density pulse in a flow with constant unitary velocity and free of source terms (i.e. a convection-dominated flow). This square test is performed on a uniform mesh ([0, 1]) with 200 cells such that the time step for the CFL condition is $\Delta t_{CFL} = 5 \times 10^{-3}$. Fig. 4(a) shows that the obtained solution is diffusive if the time step is smaller than $\Delta t_{CFL}$. As the time step increases, the error in the projection step decreases. The physical interpretation of this condition is that the nodes should 'run across' the cells of the reference mesh which contain them, eliminating the diffusion effect caused by the linear interpolation. In order to illustrate the accuracy of the method using a higher-order interpolation method, the square pulse tests are repeated considering cubic interpolation instead (fig. 4(b)). Even though it is expected that a higher-order interpolation method overcomes the excessive numerical diffusion, it introduces significant oscillations around steep density gradients. For this reason, the positivity preserving linear interpolation is used for all the numerical experiments in this paper.

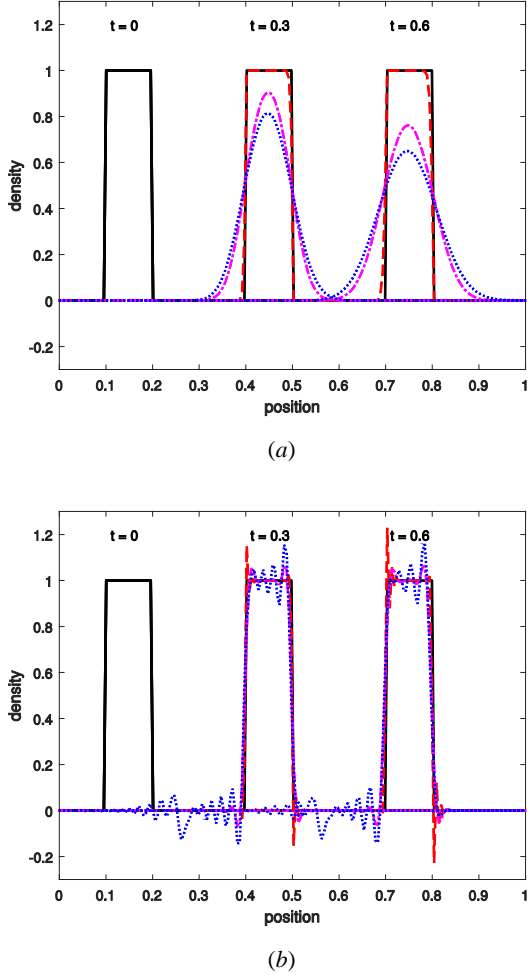

Fig. 4. Results of a square test calculated with POSS using (*a*) linear interpolation and (*b*) cubic interpolation with different time stepping: $\Delta t = 2 \times 10^{-2}$ (dashed lines), $\Delta t = 2 \times 10^{-3}$ (dashdot lines) and $\Delta t = 2 \times 10^{-4}$ (dotted lines). The exact solution is also plotted (solid lines).

Thus, the minimum time step should be chosen to avoid excessive numerical diffusion in the solution such that the displacement of each node in the auxiliary mesh is larger than the corresponding space stepping $\Delta x_i$ of the reference mesh. This condition defines that the time step $\Delta t$ in the simulation should be larger than the lower bound of accuracy (LBA) of the step time $\Delta t_{LBA}$, defined as:

$$\Delta t_{LBA}(t) = \max_{1 \leq i \leq n}\left(\left|\frac{\Delta x_i}{u_i}\right|\right) \quad (12)$$

It is interesting to compare $\Delta t_{LBA}$ with the time step restricted by the CFL condition expressed as

$$\Delta t_{CFL}(t) = \min_{1 \leq i \leq n}\left(\left|\frac{\Delta x_i}{u_i}\right|\right) \quad (13)$$

Observe that $\Delta t_{UBS}$ is much larger than $\Delta t_{LBA}$ in most cases, making POSS faster than other methods. If both UBS and LBA condition are satisfied, the error of POSS is mainly introduced by the projection step. It is vital and necessary to point out that in some cases $\Delta t_{UBS}$ can conflict with $\Delta t_{LBA}$. For example, local values of $\frac{\Delta x_i}{u_i}$ in a low-velocity region can be very large, resulting in a large $\Delta t_{LBA}$. In such a situation, the UBS condition prevails over the LBA condition. Then a certain degree of numerical diffusion will be caused at low-velocity regions due to the projection error. However, the numerical experiments in Section V show that the introduced numerical diffusion is rather small since the density $\rho$ and velocity field $\boldsymbol{u}$ in these places are always non-stiff. For this reason, POSS could ensure high-resolution even though the LBA condition is not satisfied.

For POSS, both the time splitting and interpolation can break up the mass conservation. Being consistent with the computation error expressed as (11), the mass conservation is affected significantly by $\Delta t$, which is mainly determined by the velocity field ($\boldsymbol{u}$). In later numerical experiments of this paper, the results of POSS are compared with either analytical solution or widely accepted numerical solution to demonstrate the general characteristics of mass conservation.

*C. Time complexity*

One of the most notable advantages of POSS is that it is computationally efficient. Observe that the computation time for Eulerian methods grow up as a power-law function of the number of cells in high dimensions (2D/3D). Moreover, smaller mesh size leads to a smaller time step for explicit Eulerian methods, making the whole computation time much longer. On the contrary, the computation time of POSS increases almost linearly with the number of nodes when linear interpolation is used, i.e. $t \sim nlog(n)$ since Delaunay triangulation is used. Furthermore, the time step required for POSS is almost independent of the mesh size since it weakly affects the UBS condition. Thus, the mesh size (determining the number of unknowns) of POSS could be very small to deduce the overall numerical error (see expression (11)). In higher dimensions when a large number of unknowns are introduced, the most time-consuming part is the projection step, which could be accelerated using parallel computation.

*D. Extension to other coordinate systems*

The extension of POSS to cylindrical and spherical coordinate systems is straightforward. The basic idea is to solve the position subproblem on Cartesian coordinate by replacing '$\rho$' with '$r\rho$' and '$r^2\rho$', respectively. At the same time, the state subproblem is solved in the corresponding coordinate systems.

Let us take the 2D cylindrical coordinate system (*r-z* system) as an example. Re-write equation (4) in *r-z* system below

$$\frac{\partial r\rho}{\partial t} + \left(u_r \frac{\partial r\rho}{\partial r} + u_z \frac{\partial r\rho}{\partial z}\right) = 0 \quad (14)$$

In this equation the term '$r\rho$' can be viewed as the 'modified' conserved density in a Cartesian coordinate system. This value of $\rho$ is obtained by mapping of '$r\rho$' from the auxiliary mesh to the reference mesh and then divided by $r$. However, the evaluation along the symmetry axis should be

7made with care by adding a very small tolerance to $r$ in order to avoid the division by zero.

*E. Application to corona discharges simulation*

The drift of charged particles in corona discharges is caused by the electromagnetic force, which can be obtained by solving Maxwell equations. As a first approach, only electrostatic forces given by Poisson equation are considered in this paper. Thus, the velocity $\boldsymbol{u}$ in the continuity equation (1) for each charge particle is a function of the reduced field:

$$\boldsymbol{u} = F(|\boldsymbol{E}|/N) \quad (15)$$

where $N$ is the neutral gas density and $\boldsymbol{E}$ is the electric field. Of all the charged species, electrons are the most difficult particle to handle due to their higher speed. In typical simulations of corona discharges at atmospheric pressure in air, the typical length of the ionization front is about 0.01 cm. In order to resolve accurately that ionization zone, at least 10 mesh cells are required leading to a minimum mesh size $\Delta x$ equal to 10 μm. The time step satisfying the CFL condition for simulations with standard Eulerian methods $\Delta t_{CFL}$ is roughly $10^{-11}$ s. Instead, the upper bound of stability of POSS $\Delta t_{UBS}$ is about $10^{-10}$ s considering an extreme velocity drop at the front of ionization wave (assumed e.g. $\Delta u = 10^7$ cm/s). This means that the time step required by POSS would be about one order of magnitude larger than $\Delta t_{CFL}$.

However, a smaller time step would be then required if streamer-like ionization waves develop in the simulation due to the strong variations in both the velocity vector $u$ and the reaction terms in (1), caused by a significant change in local electric fields due to significant accumulation of space charge. Under those conditions, there is a strong coupling between the continuity equations and the electric fields in the geometry, which can be characterized by the effective ionization time (EIT) and the dielectric relaxation time (DRT) [32]. The restriction in the time stepping due to the EIT is given by

$$\Delta t_{EIT}(t) = \min_{1 \leq i \leq n}\left(\frac{1}{|\alpha_i - \eta_i||\boldsymbol{u}_i|}\right) \quad (16)$$

where $\alpha_i$, $\eta_i$ and $\boldsymbol{u}_i$ denote the ionization and attachment coefficients and velocity of electrons, respectively. The subscript $i$ refers to the $i^{th}$ node in the mesh. For glow corona discharges, $\alpha_i$ is close to $\eta_i$, resulting in $\Delta t_{EIT}$ larger than $\Delta t_{UBS}$. However, as the transition to streamer corona discharges take place, $\alpha_i$ increases by several orders of magnitudes (as $\eta_i$ decreases), leading to a rapidly decreasing $\Delta t_{EIT}$. Under such condition, the simulation time stepping is restricted by $\Delta t_{EIT}$ as it becomes smaller than $\Delta t_{UBS}$.

The other physical restriction of the time step due to the dielectric relaxation is expressed as

$$\Delta t_{DRT}(t) = \min_{1 \leq i \leq n}\left(\frac{\varepsilon}{e\mu_{e,i}\rho_{e,i}}\right) \quad (17)$$

where $\varepsilon$ the permittivity of gas medium and $\mu_e$ is the electron mobility. For glow corona discharges at atmospheric pressure in air, typical maximum values of $\mu_e$ and $\rho_e$ are 400 cm$^2$/Vs and $10^{12}$ cm$^{-3}$ [33], which results in $\Delta t_{DRT} \approx 10^{-9}$s. As the electron density $\rho_e$ increases due to the transition to streamer-like structures, the dielectric relaxation time $\Delta t_{DRT}$ can decrease to values significantly smaller than $\Delta t_{CFL}$.

In the simulation of corona discharges, the numerical restriction ($\Delta t_{UBS}$) and the physical restrictions ($\Delta t_{EIT}, \Delta t_{DRT}$) have to be taken into account. As suggested in [32], the time step of the simulation is here calculated as

$$\Delta t = \min(A_E \Delta t_{EIT}, A_D \Delta t_{DRT}, A_U \Delta t_{UBS}) \quad (18)$$

with $A_E = A_D = A_U = 0.5$.

## V. NUMERICAL EXPERIMENTS

*A. General advection-diffusion problem test*

There exists analytical solutions for the 1D advection-diffusion equation even with variable coefficients [34], which are suitable for comparison with the numerical solution with POSS. For example, let us consider the problem for $D = (1-x)^2, u = 1-x$ and $f = 0$ in equation (1) with the boundary condition

$$\rho(0,t) = 1, \rho(1,t) = 0 \quad (19)$$

The exact solution is

$$\rho(x,t) = \frac{1}{1-x}\operatorname{erfc}\left\{\frac{\ln(1-x)}{-2\sqrt{t}}\right\} \quad (20)$$

where erfc is the complementary error function.

Note that in this case, the diffusion coefficient is not constant. Thus equation (4) turns into

$$\frac{\partial \rho}{\partial t} + \left(u - \frac{\partial D}{\partial x}\right)\frac{\partial \rho}{\partial x} = 0 \quad (21)$$

which is solved by first order Euler method. The equation (3) is solved by first order BDF method. The calculations are performed for a space step $5 \times 10^{-3}$ and a time step of $10^{-3}$. Here a small time step is used since this problem is not convection-dominated. Fig. 5 shows the comparison between the analytic solution and the numerical results obtained with POSS for this case. As it can be seen, there is an excellent agreement between the curves, validating the POSS to solve the general advection-diffusion problems with changing diffusion coefficient.

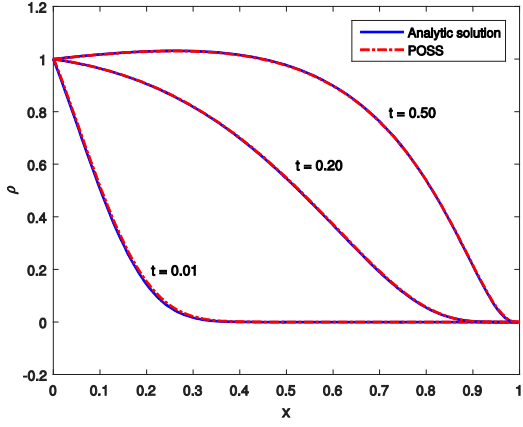

Fig. 5. Comparison between the analytic solutions and numerical results obtained with POSS.

## B. Davies test

This test was first introduced by Davies and Niessen [35] to test the performance of high-resolution algorithms for simulation of electrical discharges. The test consists of the transport of a square density profile through a stationary oscillating velocity field, as shown in fig. 6(a). The equations below give the expression for the density and velocity, respectively.

$$\rho_0(x) = \begin{cases} 10, & if\ 0.05 \leq x \leq 0.25 \\ 0, & otherwise \end{cases} \quad (22)$$

$$u(x) = 1 + 9sin^8(\pi x) \quad (23)$$

which is governed by the equation (1) with $D = 0, f = 0$.

After one period $T \approx 0.591$, the square will return back to its original position (mass conserved). This problem was solved using POSS with 400 uniform cells. The time step is set to $4.7 \times 10^{-3}$. Both the position equation and state equation are solved using the second order RK method. Fig. 6(b) shows a comparison of the solution at two different time instants calculated with the FVM-MUSCL, FEM-FCT and POSS. Observe the excellent agreement of the estimations with POSS and the analytic solution. The average of the absolute error after one period is equal to 0.2650 and 0.2677 for FVM-MUSCL and FEM-FCT algorithms, respectively [36]. This error is only 0.06 for POSS with the same mesh. The time step used in this test is about 20 times larger than $\Delta t_{CFL} = 2.5 \times 10^{-4}$.

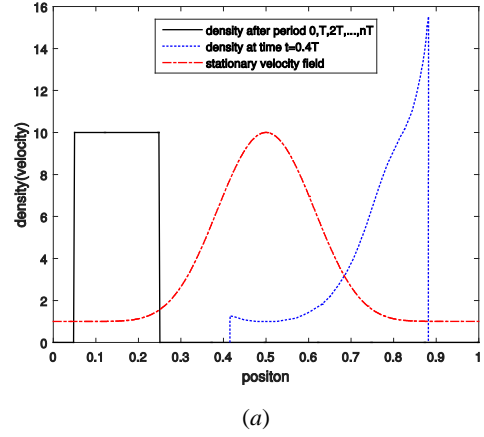

(*a*)

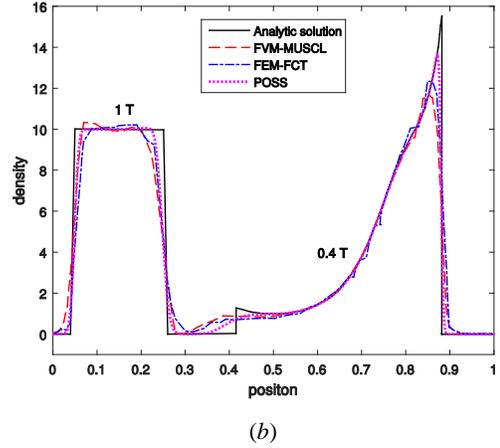

(*b*)

Fig. 6. (*a*) Initial conditions for the density and the velocity field and exact solution after 0.4 period. (*b*) Density solution at times 0.4 T and T obtained from the FVM-MUSCL, FEM-FCT algorithms and POSS.

Davies test case can be easily extended to higher dimensions. The initial condition is then given by

$$\rho_0(j_1, j_2, \dots) = \begin{cases} 10, & if\ 0.05 \leq j_1, j_2, \dots \leq 0.25 \\ 0, & otherwise \end{cases} \quad (24)$$

where $j_1, j_2, \dots$ represent dimensions. The velocity is then defined by

$$u_j = 1 + 9sin^8(\pi j), \qquad j = j_1, j_2, \dots \quad (25)$$

Davies tests in 2D and 3D are also performed to validate POSS in higher dimensions (although the results are not shown here). Excellent agreement with the analytical solution is also obtained. Furthermore, 2D Davies tests show that the computation time of using POSS is roughly linear to the number of unknowns, as shown in fig. 7. Since the projection step dominates the execution time of the simulation, POSS has linear complexity when linear interpolation is used.

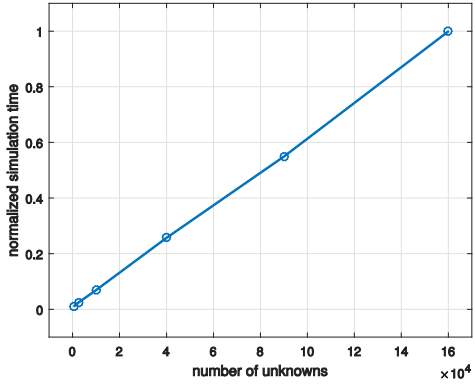

Fig. 7. The simulation time of 2D Davies test as a function of the number of unknowns, normalized to the maximum execution time.

### C. Simulation of a 1D corona discharges

In the previous two tests in this section, the velocity field in the calculations is stationary and does not vary significantly in space. Under such conditions, POSS is both accurate and computationally efficient. Now let us evaluate POSS for the simulation of a glow corona discharge, where the electric field changes weakly both in space and time.

As an example, the positive glow corona simulation in a 1D coaxial spherical configuration presented in [33] is here performed using POSS. The set of continuity equations for electrons, positive ions, negative ions and metastable molecules in spherical coordinates [33] is given by

$$\frac{\partial N_e}{\partial t} = S + (\alpha - \eta)N_e|\boldsymbol{W}_e| - \beta N_e N_p \\ + k_d O_2^* O_2^- - \frac{1}{r^2}\frac{\partial(r^2 N_e \boldsymbol{W}_e)}{\partial r} \quad (26)$$

$$\frac{\partial N_p}{\partial t} = S + \alpha N_e|\boldsymbol{W}_e| - \beta N_e N_p - \beta N_p O_2^- \\ - \frac{1}{r^2}\frac{\partial(r^2 N_p \boldsymbol{W}_p)}{\partial r} \quad (27)$$

$$\frac{\partial O_2^-}{\partial t} = \eta N_e|\boldsymbol{W}_e| - k_d O_2^* O_2^- - \beta N_p O_2^- \\ - \frac{1}{r^2}\frac{\partial(r^2 O_2^- \boldsymbol{W}_n)}{\partial r} \quad (28)$$

$$\frac{\partial O_2^*}{\partial t} = \alpha_m N_e|\boldsymbol{W}_e| - k_d O_2^* O_2^- - k_q O_2^* O_2 \quad (29)$$

where $t$ is the time, $N_e, N_p, O_2^-, O_2$ and $O_2^*$ are the number densities of electrons, positive ions, negative ions, oxygen molecules and metastable ($^1a\Delta_g$) oxygen molecules, respectively. $\boldsymbol{W}_e, \boldsymbol{W}_p, \boldsymbol{W}_n$ are the drift velocities for electrons, positive ions and negative ions. The gas medium is air at atmospheric pressure and the symbols $\alpha, \eta, \beta, \alpha_m$ denote the ionization, attachment, electron-ion (ion-ion) recombination coefficients and the rate of creation of metastable oxygen molecules due to electron impact, respectively. $k_d, k_q$ are the detachment rate coefficient and quenching rate constant, respectively. The diffusion movement of charged particles is neglected. $S$ is the photo-ionization source term.

In the simulation, a voltage of 20 kV is applied to the inner conductor (radius 0.1cm) while the outer conductor (radius 2.1cm) is grounded. The computation domain is discretized into a non-uniform mesh with 800 cells. The mesh size increases exponentially from the inner conductor surface to outer conductor surface. The material functions for air and the model for photo-ionization are exactly the same as in [33].

The time step $\Delta t$ in the simulation changes dynamically according to equation (18). All the position and state subproblems are solved using the first order Euler method. Fig. 8(*a*) and (b) plots the electric filed at different instants and the total corona discharge current, respectively. An excellent agreement with the estimates in [33] is obtained.

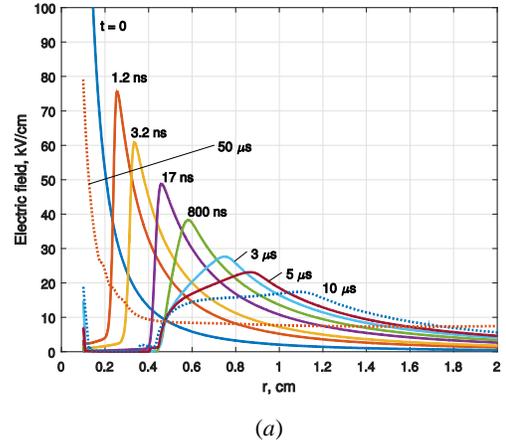

(*a*)

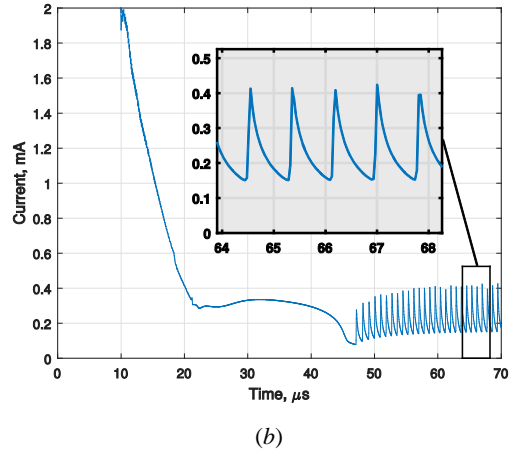

(*b*)

Fig. 8. (*a*) Electric field distribution at different time instants. (*b*) The corona discharge current.

Fig. 9 illustrates the time step used in the simulation as well as $\Delta t_{UBS}, \Delta t_{EIT}$ and $\Delta t_{CFL}$. $\Delta t_{DRT}$ is not shown since it is much larger than other time step restrictions. Overall, the time step used in this simulation is 5~20 times larger than $\Delta t_{CFL}$, which is the main restriction for other Eulerian methods for example the FD-FCT method used in [33]. Thus the time step of POSS is significantly larger than for FD-FCT when simulating glow corona discharges. Note that such advantage will not exist for streamer discharge simulation where electric field changes dramatically with time and space. For streamer simulation, $\Delta t$ is significantly smaller than $\Delta t_{CFL}$. The mass will not be conserved due to excessive diffusion mainly caused by interpolation.

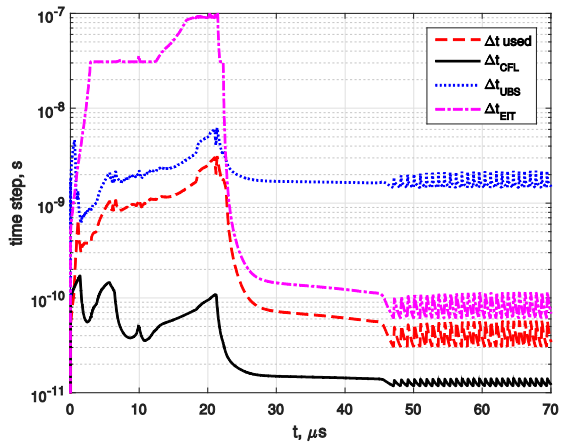

Fig. 9. The time step used in the simulation compared with $\Delta t_{CFL}$, $\Delta t_{UBS}$ and $\Delta t_{EIT}$.

The extension of POSS to simulate corona discharges in higher dimensions is straightforward. The related simulation results will be reported in the forthcoming papers.

## VI. Conclusions

An efficient semi-Lagrangian algorithm for simulation of corona discharges is proposed. Several numerical experiments are conducted to demonstrate the low computational cost, robustness, and high-resolution of the position-state separation method (POSS) to solve convection-dominated continuity equations. Several conclusions can be drawn:

- For the simulation of corona discharges where the velocity field is weakly changing in time, the solution with POSS is not restricted by the CFL condition when solving the continuity equations. Therefore, a time step significantly larger than that for explicit Eulerian methods can be used. The time complexity of POSS is linear to the unknowns when dealing with the convection-dominated flows.

- As a semi-Lagrangian method, the overall error of POSS can be reduced by using a suitable time step lower than the upper bond of stability $\Delta t_{UBS}$. In addition, even linear interpolation can ensure high resolution as long as the time step is larger than the lower bond of accuracy $\Delta t_{LBA}$. POSS can also be easily extended to cylindrical and other coordinate systems and high dimensions.

- Without flux correction and combined with a finite element solver for the state subproblem, POSS is easy to be implemented on arbitrary geometries.

These features enable POSS to have great potential in modeling of corona discharges in large interelectrode gaps for long simulation times.

Acknowledgment

LL greatly appreciates the scholarship support from China Scholarship Council (Grant No. 201306160003). MB would like to acknowledge the financial support of the Swedish strategic research program StandUp for Energy.

References

[1] Y. P. Raizer, *et al.*, *Gas discharge physics*: Springer-Verlag Berlin, 1991.
[2] R. Waters and W. Stark, "Characteristics of the stabilized glow discharge in air," *Journal of Physics D: Applied Physics,* vol. 8, p. 416, 1975.
[3] S. Dhali and P. Williams, "Two dimensional studies of streamers in gases," *Journal of applied physics,* vol. 62, pp. 4696-4707, 1987.
[4] J. P. Boris and D. L. Book, "Flux-corrected transport. I. SHASTA, A fluid transport algorithm that works," *Journal of Computational Physics,* vol. 11, pp. 38-69, 1973.
[5] D. L. Book, *et al.*, "Flux-corrected transport II: Generalizations of the method," *Journal of Computational Physics,* vol. 18, pp. 248-283, 1975.
[6] J. P. Boris and D. L. Book, "Flux-corrected transport. III. Minimal-error FCT algorithms," *Journal of Computational Physics,* vol. 20, pp. 397-431, 1976.
[7] S. T. Zalesak, "Fully multidimensional flux-corrected transport algorithms for fluids," *Journal of Computational Physics,* vol. 31, pp. 335-362, 1979.
[8] R. Morrow and L. Cram, "Flux-corrected transport and diffusion on a non-uniform mesh," *Journal of Computational Physics,* vol. 57, pp. 129-136, 1985.
[9] R. Löhner, *et al.*, "Finite element flux-corrected transport (FEM–FCT) for the euler and Navier–Stokes equations," *International Journal for Numerical Methods in Fluids,* vol. 7, pp. 1093-1109, 1987.
[10] R. Löhner, *et al.*, "A finite element solver for axisymmetric compressible flows," presented at the AIAA, Fluid Dynamics, 20th Plasma Dynamics and Lasers Conference, 1989.
[11] D. Kuzmin and S. Turek, "Flux correction tools for finite elements," *Journal of Computational Physics,* vol. 175, pp. 525-558, 2002.
[12] B. Van Leer, "Towards the ultimate conservative difference scheme. V. A second-order sequel to Godunov's method," *Journal of Computational Physics,* vol. 32, pp. 101-136, 1979.
[13] B. P. Leonard, "A stable and accurate convective modelling procedure based on quadratic upstream interpolation," *Computer Methods in Applied Mechanics and Engineering,* vol. 19, pp. 59-98, 1979.
[14] C.-W. Shu, "Total-variation-diminishing time discretizations," *SIAM Journal on Scientific and Statistical Computing,* vol. 9, pp. 1073-1084, 1988.
[15] B. Cockburn, *et al.*, "TVB Runge-Kutta local projection discontinuous Galerkin finite element method for conservation laws III: one-dimensional

systems," *Journal of Computational Physics,* vol. 84, pp. 90-113, 1989.
[16] B. Cockburn and C.-W. Shu, "TVB Runge-Kutta local projection discontinuous Galerkin finite element method for conservation laws. II. General framework," *Mathematics of Computation,* vol. 52, pp. 411-435, 1989.
[17] B. Cockburn*, et al.*, "The Runge-Kutta local projection discontinuous Galerkin finite element method for conservation laws. IV. The multidimensional case," *Mathematics of Computation,* vol. 54, pp. 545-581, 1990.
[18] B. Cockburn and C.-W. Shu, "The Runge–Kutta discontinuous Galerkin method for conservation laws V: multidimensional systems," *Journal of Computational Physics,* vol. 141, pp. 199-224, 1998.
[19] C. Johnson, "A new approach to algorithms for convection problems which are based on exact transport + projection," *Computer Methods in Applied Mechanics and Engineering,* vol. 100, pp. 45-62, 1992.
[20] O. Shipilova*, et al.*, "Particle transport method for convection problems with reaction and diffusion," *International Journal for Numerical Methods in Fluids,* vol. 54, pp. 1215-1238, 2007.
[21] A. Smolianski*, et al.*, "A fast high‐resolution algorithm for linear convection problems: particle transport method," *International Journal for Numerical Methods in Engineering,* vol. 70, pp. 655-684, 2007.
[22] O. Pironneau*, et al.*, "Characteristic-Galerkin and Galerkin/least-squares space-time formulations for the advection-diffusion equation with time-dependent domains," *Computer Methods in Applied Mechanics and Engineering,* vol. 100, pp. 117-141, 1992.
[23] A. Robert, "A stable numerical integration scheme for the primitive meteorological equations," *Atmosphere-Ocean,* vol. 19, pp. 35-46, 1981.
[24] D. Xiu and G. E. Karniadakis, "A Semi-Lagrangian High-Order Method for Navier–Stokes Equations," *Journal of Computational Physics,* vol. 172, pp. 658-684, 2001.
[25] O. Chanrion and T. Neubert, "A PIC-MCC code for simulation of streamer propagation in air," *Journal of Computational Physics,* vol. 227, pp. 7222-7245, 2008.
[26] H. Okuda, "Nonphysical noises and instabilities in plasma simulation due to a spatial grid," *Journal of Computational Physics,* vol. 10, pp. 475-486, 1972.
[27] G. I. Marchuk, "Splitting and alternating direction methods," in *Handbook of Numerical Analysis*. vol. Volume 1, P. G. Ciarlet and J. L. Lions, Eds.: Elsevier, 1990, pp. 197-462.
[28] J. Alberty*, et al.*, "Remarks around 50 lines of Matlab: short finite element implementation," *Numerical Algorithms,* vol. 20, pp. 117-137, 1999.
[29] R. W. Lewis*, et al.*, *Fundamentals of the finite element method for heat and fluid flow*: John Wiley & Sons, Ltd, 2004.
[30] D. S. Kershaw, "The incomplete Cholesky—conjugate gradient method for the iterative solution of systems of linear equations," *Journal of Computational Physics,* vol. 26, pp. 43-65, 1978.
[31] M. Falcone and R. Ferretti, "Convergence analysis for a class of high-order semi-Lagrangian advection schemes," *SIAM Journal on Numerical Analysis,* vol. 35, pp. 909-940, 1998.
[32] P. A. Vitello*, et al.*, "Simulation of negative-streamer dynamics in nitrogen," *Physical Review E,* vol. 49, pp. 5574-5598, 1994.
[33] R. Morrow, "The theory of positive glow corona," *Journal of Physics D: Applied Physics,* vol. 30, p. 3099, 1997.
[34] A. Kumar*, et al.*, "Analytical solutions to one-dimensional advection–diffusion equation with variable coefficients in semi-infinite media," *Journal of Hydrology,* vol. 380, pp. 330-337, 2010.
[35] A. Davies and W. Niessen, "The solution of the continuity equations in Ionization and plasma growth," in *Physics and Applications of Pseudosparks*: Springer, 1990, pp. 197-217.
[36] O. Ducasse*, et al.*, "Critical analysis on two-dimensional point-to-plane streamer simulations using the finite element and finite volume methods," *Plasma Science, IEEE Transactions on,* vol. 35, pp. 1287-1300, 2007.